\documentclass[twocolumn,aps,floatfix,prl,superscriptaddress]{revtex4}
\usepackage{amsmath,amssymb,eucal,graphicx,float}
\begin{document}
\title{Kinetics of Diffusion-Controlled Annihilation 
with Sparse Initial Conditions} 
\author{E.~Ben-Naim}
\affiliation{Theoretical Division and Center for Nonlinear Studies, 
Los Alamos National Laboratory, Los Alamos, New Mexico 87545, USA} 
\author{P.~L.~Krapivsky}
\affiliation{Department of Physics, Boston University, Boston,
  Massachusetts 02215, USA}
\begin{abstract}
We study diffusion-controlled single-species annihilation with sparse
initial conditions. In this random process, particles undergo Brownian
motion, and when two particles meet, both disappear. We focus on
sparse initial conditions where particles occupy a subspace of
dimension $\delta$ that is embedded in a larger space of dimension
$d$. We find that the co-dimension $\Delta=d-\delta$ governs the
behavior. All particles disappear when the co-dimension is 
sufficiently small, $\Delta\leq 2$; otherwise, a finite fraction of
particles indefinitely survive.  We establish the asymptotic behavior
of the probability $S(t)$ that a test particle survives until time
$t$.  When the subspace is a line, $\delta=1$, we find inverse
logarithmic decay, $S\sim (\ln t)^{-1}$, in three dimensions, and a
modified power-law decay, $S\sim (\ln t)\,t^{-1/2}$, in two
dimensions. In general, the survival probability decays algebraically
when $\Delta <2$, and there is an inverse logarithmic decay at the
critical co-dimension $\Delta=2$.
\end{abstract}
%\pacs{05.40.--a, 82.20.--w, 66.10.C--,  05.70.Ln}     
\maketitle

Reaction-diffusion processes are ubiquitous in a wide range of natural
and physical phenomena \cite{pg,skf,otb}. Minimal models of
diffusion-limited aggregation, coalescence, and annihilation play a
central role in the theory of fractals \cite{ws}, pattern formation
\cite{cg,vvs}, and non-equilibrium statistical physics \cite{thv}. In
the latter context, the central finding is that in low dimensions,
spatial correlations dominate and lead to slow reaction kinetics. This
behavior has been observed experimentally \cite{kfs,rmkrtl,addlwsylb},
and it well understood theoretically
\cite{zo,bg,tw,pgg,kr,tm,jls,brt,soo,bbd,db,mb}.

The single-species annihilation process has played a key role in
illuminating the statistical mechanics of reaction kinetics.  It is
represented by the reaction scheme
\begin{equation}
\label{process}
A+A\to \emptyset\,.
\end{equation}
In this random process, particles undergo ordinary diffusion and
whenever two particles meet, they both disappear.  Starting with a
uniform density, the survival probability of a test particle, $S(t)$,
exhibits the long-time asymptotic behaviors \cite{tm,kr,bg}
\begin{equation}
\label{uniform}
S(t)\sim 
\begin{cases}
t^{-d/2}          & d<2;\\
(\ln t)\,t^{-1}  & d=2;\\
t^{-1}            & d>2. 
\end{cases}
\end{equation}
The decay is slow below the critical dimension, while in the
complementary case, there is a quick decay, and the exponent does not
depend on the dimension. Strong spatial correlations in the positions
of the surviving particles, which develop below the critical dimension
$d_c=2$, are responsible for this behavior \cite{bg,bh,krb}.

Previous studies of single-species diffusion-controlled annihilation
have mostly dealt with spatially-homogeneous situations where the
initial density is uniform.  Among a few exceptions are the analyses
of single-species reaction processes in one dimension where particles
initially occupy the half-line \cite{bbd,db,fkr,kb}.  There are also a
few studies of single-species annihilation with a finite (but large)
number of particles \cite{kpwh,fds}. A recent study shows that the
survival probability is altered dramatically when the initial number
of particles is finite, as a finite {\it number} of the particles may
survive indefinitely \cite{bk}.

\begin{figure}
\centering
\includegraphics[width=6cm]{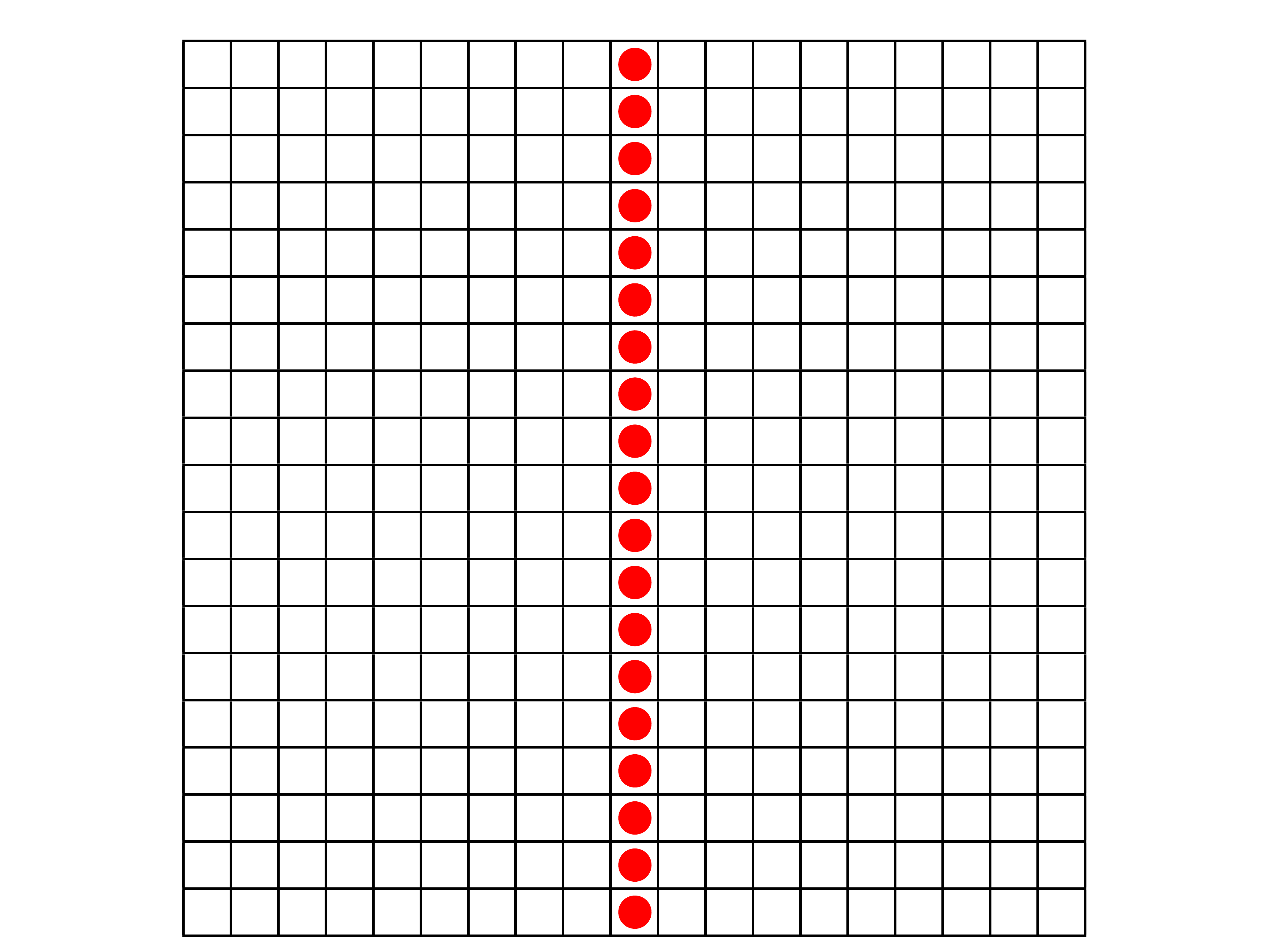}
\caption{An initial conditions where particles, denoted by red dots,
  occupy a line in two dimensions ($\delta=1$, $d=2$).}
\label{Fig:illust}
\end{figure}

Motivated by these results, we consider reaction kinetics of
single-species annihilation with another sparse initial condition.
Specifically, we consider beginning configurations where particles
occupy a subspace with dimension $\delta$ that is embedded in a space
with larger dimension $d$.  We consider the simplest possible case
where the initial distribution of particles is uniform inside the
occupied sub-space (figure \ref{Fig:illust} illustrates a line in two
dimensions).

Our main result is that the co-dimension $\Delta=d-\delta$ controls
the asymptotic behavior.  For dimensions $d>2$, we obtain
\begin{equation}
\label{main}
S(t)\sim 
\begin{cases}
t^{-(2-\Delta)/2}                                              & \Delta<2,\\
(\ln t)^{-1}                                                 & \Delta=2,\\
S_\infty +{\rm const.}\times t^{-(\Delta-2)/2}   & \Delta>2. 
\end{cases}
\end{equation}
All particles disappear when the co-dimension is sufficiently small,
$\Delta<2$, and the survival probability decays as a power law in this
case. A finite {\it fraction} $S_\infty\equiv S(\infty)$ of the
particles survives indefinitely when the co-dimension is large enough,
$\Delta>2$.  Finally, all particles disappear in the borderline case
$\Delta=2$, although the survival probability decays extremely slowly,
as inverse the logarithm of time.

The single-species annihilation process is tractable analytically only
in one dimension (see e.g. \cite{tm,mb,bh,krb,kb}). Therefore our
arguments in favor of \eqref{main} rely on heuristic reasoning.  We
start with three dimensions, $d=3$, and derive \eqref{main} for
initial conditions where particles initially occupy a line,
$\delta=1$, or a plane, $\delta=2$. We then analyze a line,
$\delta=1$, embedded in a two-dimensional space, $d=2$.  We use
scaling methods to obtain \eqref{main} from the reaction-diffusion
equation, and thereby adopt a similar approach to that used to analyze
initial conditions with a finite number of particles \cite{bk}.

In single-species annihilation, identical particles undergo diffusion
and whenever two particles touch each other, both particles
disappear. This process can be realized in discrete or in continuum
space.  In the discrete version, particles move in an unbounded
hyper-cubic lattice of dimension $d$, and each lattice site can be
occupied by at most one particle. Particles move asynchronously and
independently. In each hoping event, a particle moves to one of its
$2d$ neighboring sites, chosen at random.  If the particle lands on an
occupied site, both particles disappear. Initially, the particles
occupy a subspace of dimension $\delta$.  We implemented this discrete
version in our numerical simulations.

We begin with the three-dimensional case, $d=3$, and our starting
point is the standard reaction-diffusion equation for the density
$c\equiv c(x,y,z,t)$
\begin{equation}
\label{rde}
\frac{\partial c}{\partial t} = D\nabla^2 c - K\,c^2 \,.
\end{equation}
Here $D$ is the diffusion coefficient, $\nabla^2$ the Laplace
operator, and $K$ the reaction coefficient. The reaction term is
quadratic in $c$ reflecting that the reaction process \eqref{process}
involves two particles \cite{mvs}. For a spatially uniform system, the
diffusion term vanishes, so \eqref{rde} reduces to $dc/dt=-Kc^2$, and
the density decay $c\sim t^{-1}$ in \eqref{uniform} follows.  Of
course, for homogeneous systems, the survival probability of a test
particle is proportional to the density, $S\propto c$. Here, and in
the rest of this study, we set the diffusion coefficient and the
reaction coefficient to unity $D=K=1$, as we are primarily interested
in asymptotic behaviors.

We now treat the case where the particles occupy a line,
$\delta=1$. Without loss of generality we set this line as the
$z$-axis in a standard Cartesian coordinate system. With this setup,
we expect that $\partial_z c=0$. Further, the Laplace operator in
\eqref{rde} becomes two-dimensional, \hbox{$\nabla^2\equiv
  \partial_x^2 + \partial_y^2$}.  The density $c\equiv c(x,y,t)$
depends only on two of the three coordinates, and the survival
probability of a particle is the integrated density
\begin{equation}
\label{S31-def}
S(t)=\iint dx\,dy\,c(x,y, t)\,.
\end{equation}
We stress that the survival probability equals the total number of
particles per unit length along the $z$-axis. By substituting
\eqref{S31-def} into the reaction-diffusion equation \eqref{rde}, we
find that the survival probability obeys 
\begin{equation}
\label{S31-eq}
\frac{dS}{dt} = - \iint dx\,dy\, c^2\,.
\end{equation}
The right-hand side of this equation is simply the total reaction
rate per unit length. 

\begin{figure}
\centering
\includegraphics[width=7cm]{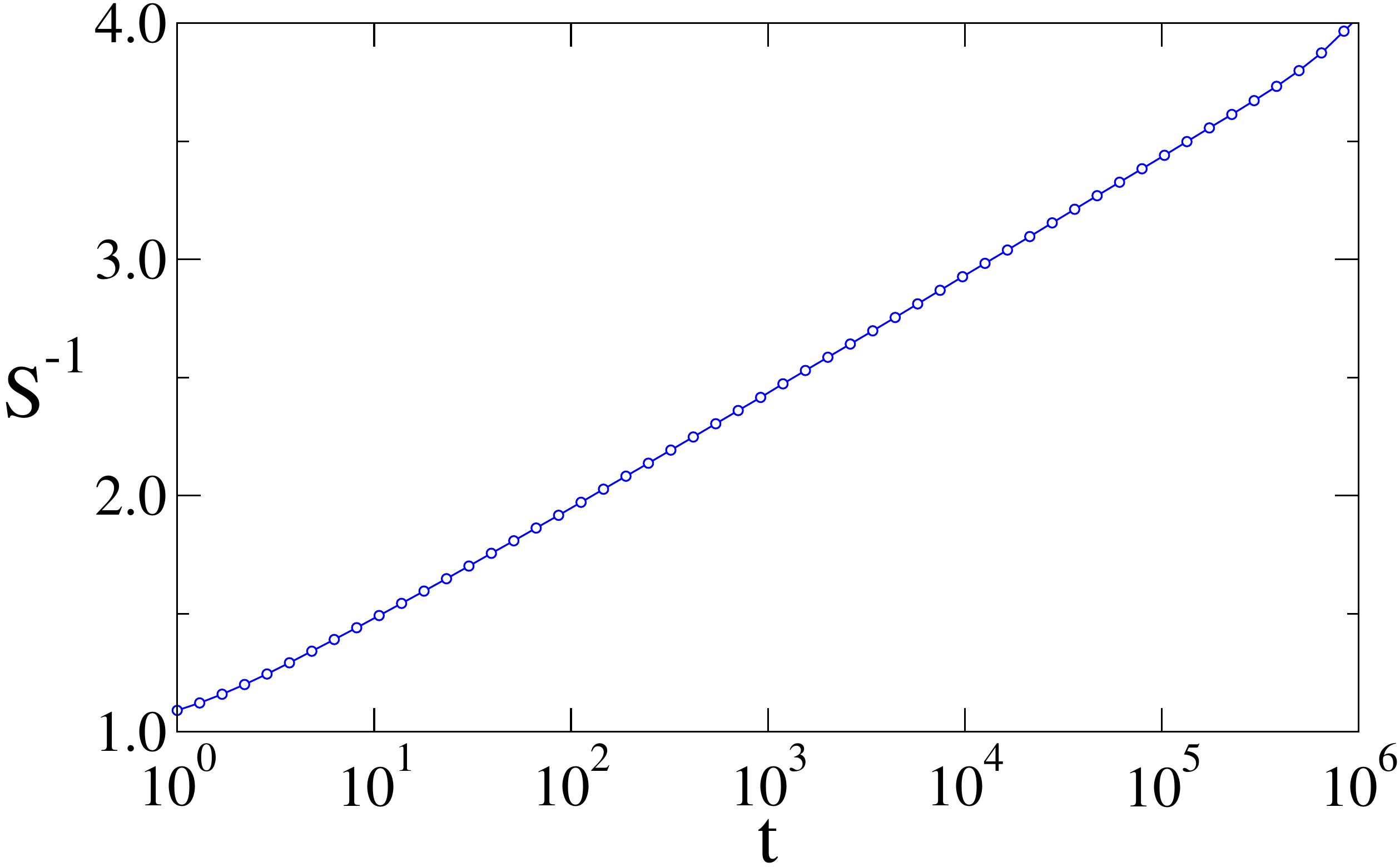}
\caption{The inverse survival probability $S^{-1}$ versus time $t$ for
  single-species annihilation in three dimension, starting with a line
  of particles ($\delta=1$).  The simulations were performed on a
  cubic lattice with $2500^3$ sites and periodic boundary
  conditions. Initially, there are $2500$ particles.}
\label{Fig31}
\end{figure}

We now use heuristic methods to estimate the total reaction rate in
\eqref{S31-eq}. The initial arrangement of the particles is
radially-symmetric along the $z$-axis, and this symmetry is maintained
throughout the reaction-diffusion process. Therefore, the density
depends only on the radial coordinate, $c\equiv c(r,t)$ with
$r=\sqrt{x^2+y^2}$. Since particles move diffusively, we expect that
particles are essentially confined to within a cylinder of growing
radius $r\sim \sqrt{t}$ whose axis coincides with the $z$-axis. Hence,
we approximate the density by
\begin{equation}
\label{crt}
c(r,t) \sim \frac{S(t)}{t} \times
\begin{cases}
1             & \quad r< \sqrt{t}\,;\\
0             & \quad r> \sqrt{t}\,.
\end{cases}
\end{equation}
In this approximation, the density is uniform inside the growing
cylinder and vanishes outside of it. By substituting the approximate
density \eqref{crt} into \eqref{S31-eq}, we obtain a rate equation for
the survival probability,
\begin{equation}
\label{S31-re}
\frac{dS}{dt} \sim -\frac{S^2}{t}\,.
\end{equation}
In the long-time limit, the survival probability is inversely
proportional to the logarithm of time, 
\begin{equation}
\label{S31}
S \sim (\ln t)^{-1}\,.
\end{equation}
This extremely slow decay demonstrates how sparse initial condition
can result in a slow reaction process.  The logarithmic time
dependence \eqref{S31} is difficult to confirm numerically, but
results of our large-scale simulations using a cubic lattice with
$2500^3=1.5625\cdot 10^{10}$ sites support the theoretical prediction
(Fig.~\ref{Fig31}).

\begin{figure}
\centering
\includegraphics[width=7cm]{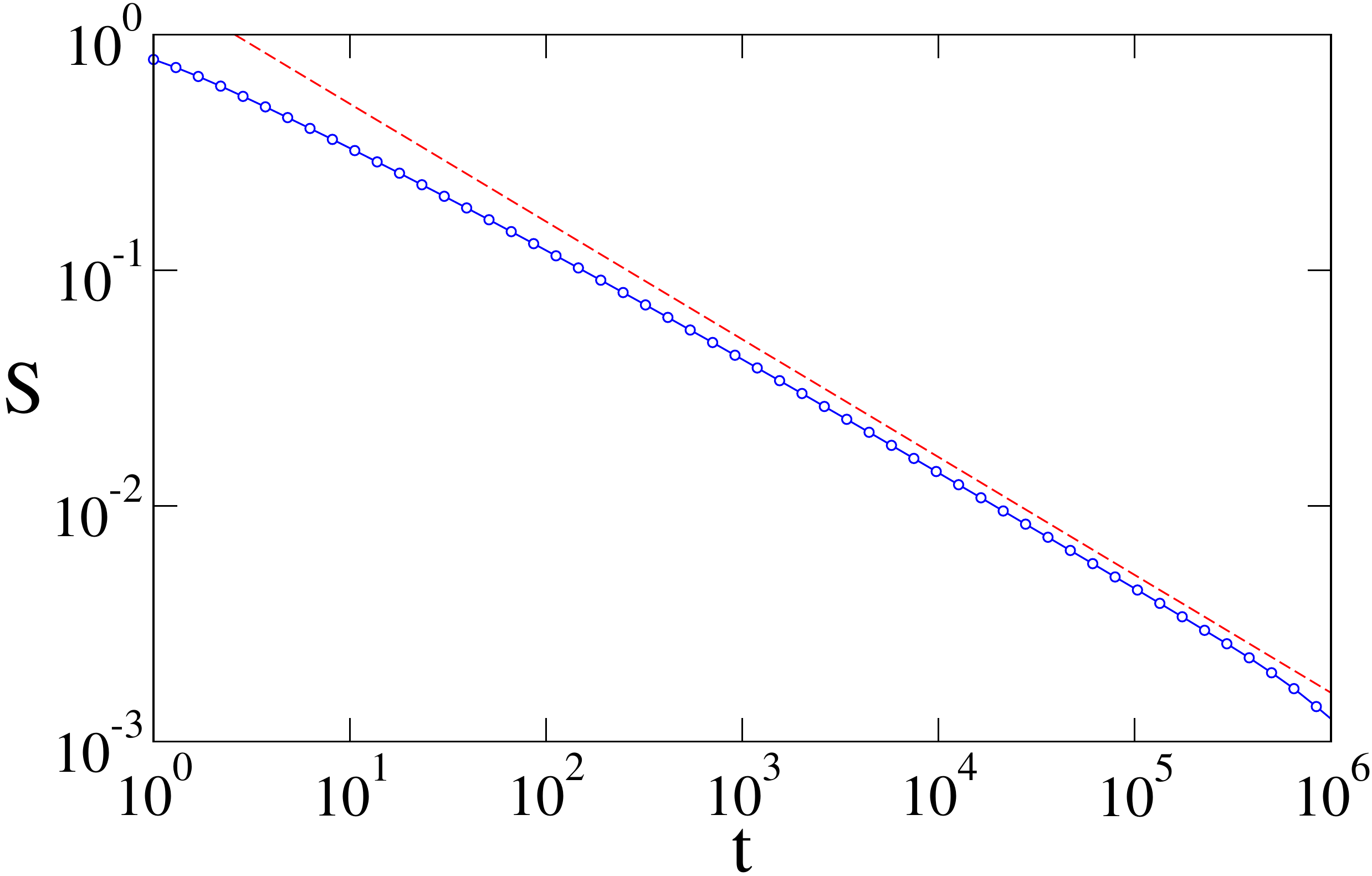}
\caption{The survival probability $S(t)$ versus time $t$ for $d=3$ and
  $\delta=2$.  The simulations were performed using a cubic lattice
  with $2500^3$ sites and periodic boundary conditions. Initially,
  $2500^2$ particles occupy a plane.}
\label{Fig32}
\end{figure}

We now consider a plane, $\delta=2$, embedded in three-dimensional
space.  The co-dimension now equals unity, \hbox{$\Delta=d-\delta=1$},
and the Laplace operator is effectively one-dimensional,
$\nabla^2\equiv \partial_x^2$.  By repeating the steps leading to
\eqref{S31-re}, we obtain
\begin{equation}
\label{S32-eq}
\frac{dS}{dt} \sim -\frac{S^2}{\sqrt{t}}.
\end{equation}
Therefore, the survival probability has the power-law decay, $S\sim
t^{-1/2}$ as stated in \eqref{main}. Results of our numerical
simulations are in good agreement with this theoretical prediction
(Fig.~\ref{Fig32}).

Equations \eqref{S31-eq} and \eqref{S32-eq} can be easily generalized
to arbitrary co-dimension $\Delta$. The general rate equation reads
$dS/dt\sim -S^2/t^{\Delta/2}$ and equation \eqref{main} follows. When
the co-dimension is large enough, $\Delta>2$, the survival probability
saturates at a constant value and consequently, a test particle has a
finite chance of avoiding the annihilation process. In the
complementary case, $\Delta < 2$, the density decays as a power law
indefinitely.

Next, we discuss the behavior at the critical dimension, $d=2$.  When
$d>2$, the reaction rate in \eqref{rde} is a constant, while for $d<2$
the reaction coefficient scales as a power of the density,
\hbox{$K\sim c^{(2-d)/d}$} \cite{krb}. At the critical dimension, the
effective reaction coefficient decays as a logarithmic function of the
density, \hbox{$K\sim 1/\ln (1/c)$}. Therefore
\begin{equation}
\label{rde2}
\frac{\partial c}{\partial t} = \nabla^2 c - \frac{c^2}{\ln(1/c)}\,.
\end{equation}
For uniform systems, the diffusion term vanishes, and equation
\eqref{rde2} simplifies to \hbox{$dc/dt\sim -c^2/\ln (1/c)$}.  The
reduced reaction rate leads to a logarithmic enhancement in the
survival probability \hbox{$\sim (\ln t)\,t^{-1}$} stated in
\eqref{uniform}.

\begin{figure}
\includegraphics[width=7cm]{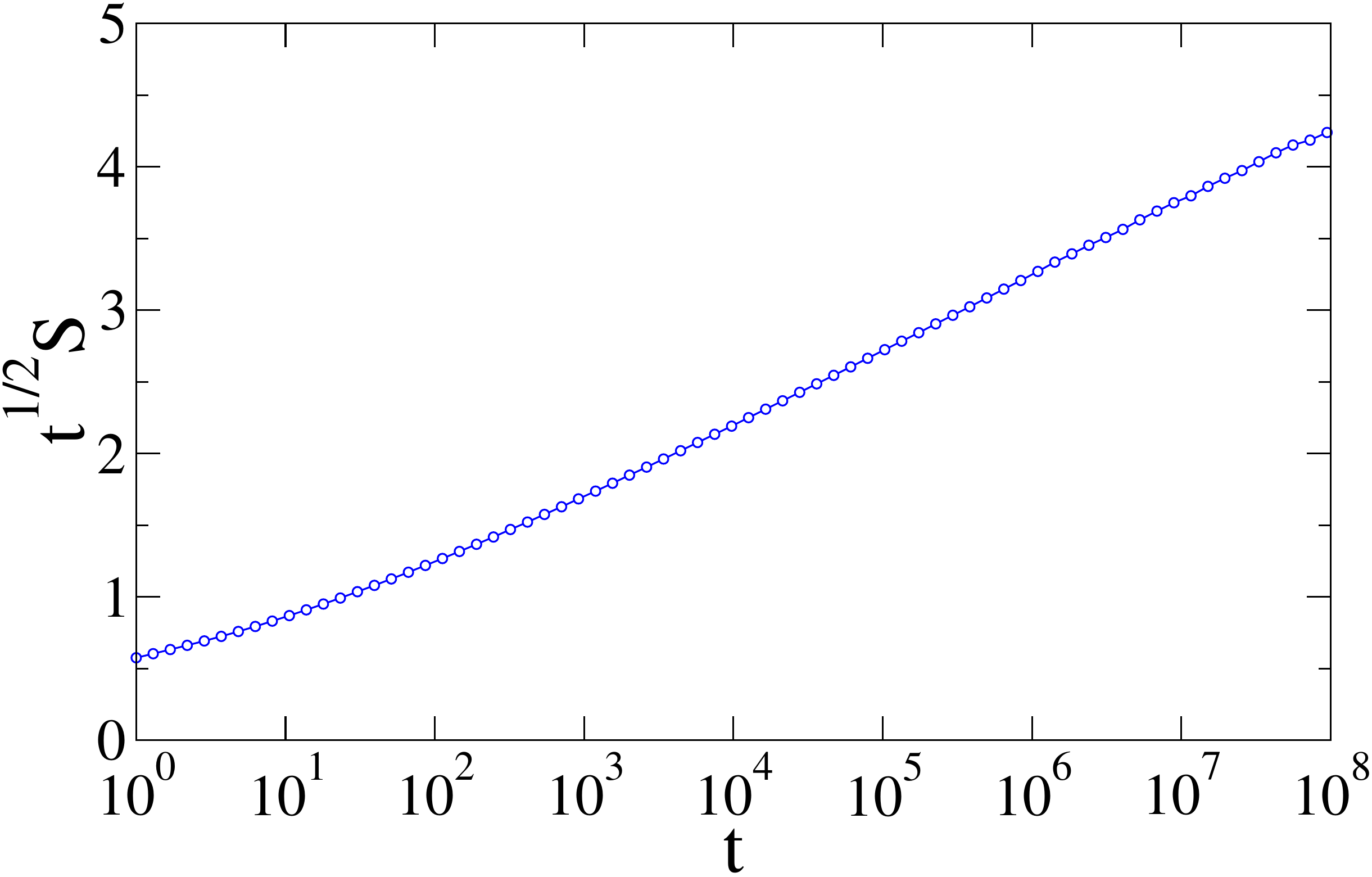}
\caption{Plot of $\sqrt{t}\,S(t)$ versus $t$ for the two-dimensional
  single-species annihilation process with initially occupied
  line. The simulations were performed on a square lattice with
  $125000^2$ sites and periodic boundary conditions, that is initially
  occupied by $125000$ particles.}
\label{Fig-21}
\end{figure}

We now turn to a line in two dimensions as illustrated in
Fig.~\ref{Fig:illust}. In a Cartesian coordinate system, we envision
the particles to occupy the $y$-axis. Now, the density is uniform
along the $y$-direction and the Laplace operator in \eqref{rde2} is
effectively one-dimensional, $\nabla^2\equiv \partial_x^2$.  By
integrating Eq.~\eqref{rde2}, we see that the survival probability,
$S(t)=\int dx\, c(x,t)$, decays according to
\begin{equation}
\label{S21-eq}
\frac{dS}{dt} = - \int dx\,  \frac{c^2}{\ln (1/c)}\,.
\end{equation}
Once again, the total reaction rate represents a rate per unit length
along a line parallel to the $y$-axis. We anticipate that the reaction
process is limited to the diffusive region \hbox{$|x|<t^{1/2}$}, and
using the same reasoning behind \eqref{S31-eq}, we obtain a rate
equation for the survival probability,
\begin{equation}
\label{S21-re}
\frac{dS}{dt} \sim - \frac{S^2}{t^{1/2}\ln (t^{1/2}/S)}\,.
\end{equation}
Therefore, the survival probability decays according to 
\begin{equation}
\label{S21}
S\sim (\ln t)\,t^{-1/2}\,.
\end{equation}
Results of our numerical simulations are in good agreement with this
prediction (Fig.~\ref{Fig-21}). 

We now discuss the complementary case where particles occupy a
sub-space with co-dimension $\Delta$ at or below the critical
dimension, \hbox{$d\leq 2$}.  In two dimensions, the effective Laplace
operator has dimension $\Delta$, and the quantity $t^{1/2}$ in the
denominator of equation \eqref{S21-re} should be replaced with
$t^{\Delta/2}$. Below the critical dimension, $d<2$, we substitute the 
effective reaction coefficient $K\sim c^{(2-d)/d}$ into \eqref{rde}.
The survival probability exhibits the
asymptotic decay
\begin{equation}
S\sim 
\begin{cases}
(\ln t)\,t^{-\delta/2}\, & d=2 ;\\
t^{-\delta/2}\,          & d<2 .\\
\end{cases}
\end{equation}  

In our numerical studies, we used a standard algorithm for simulating
the annihilation process \cite{krb}. Particles reside in a
$d$-dimensional hyper-cubic lattice with linear dimension $L$ and
periodic boundary conditions. Initially, $L^{\delta}$ particles occupy
a hyper-cubic lattice with dimension $\delta$, that is embedded in the
larger space, while the rest of the sites are empty. In each
elementary simulation step one particle, chosen at random, moves to
one of its neighboring sites. If that site is occupied, both particles
are removed from the system. Time is augmented by the inverse of the
number of remaining particles. The system memory required by this
algorithm is ${\cal O}(L^d)$, and the processing cost per unit time is
proportional to the number of surviving particles, $S(t)\times
L^\delta$.

In what follows, we show how several of the above results can be
derived using alternative probabilistic arguments.  First, we obtain
the asymptotic decay \eqref{S21} using such methods.  To derive the
$S\sim (\ln t)\,t^{-1}$ decay in \eqref{uniform}, one begins with 
the number of distinct lattice sites visited by a random walker
\cite{mp} in two dimensions, 
\begin{equation}
\label{N}
\mathcal{N}\sim \frac{t}{\ln t}\,,
\end{equation}
Then one argues that if all lattice sites are initially occupied, all
particles in a visited region coalesce into one particle, and the
density is reciprocal to this number, $S\sim \mathcal{N}^{-1}$, in
agreement with \eqref{uniform}. When the particles initially occupy a
line, we can focus on a square domain with area $\sqrt{t}\times
\sqrt{t}$. We now replace the inhomogeneous spatial distribution with
a homogeneous distribution with density $t^{-1/2}$, such that the
total initial number of particles equals the linear dimension of the
domain, $\sqrt{t}$.  The total number of particles in the visited
region scales as $t^{-1/2}\mathcal{N}$, and its reciprocal gives
\eqref{S21}.

Another key result is that when the co-dimension $\Delta>2$, each
particle has a finite probability $S_\infty>0$ of surviving the
annihilation process. This result can also be derived using
probabilistic arguments \cite{bk}, as we now demonstrate for 
a line, $\delta=1$, embedded in a $d$-dimensional
space. We consider the annihilation to take place in an unbounded
hypercubic lattice where the line $(0,\ldots,0,n)$ for all
\hbox{$-\infty<n<\infty$} is initially fully occupied while the rest
of the lattice is empty.  We also denote by $P_n$ the probability that
two random walkers, separated by distance $n$ in a $d$-dimensional
space, never meet. The probability $S_\infty$ that the particle,
initially located at the origin, survives forever has the lower bound,
\begin{equation}
\label{product}
S_\infty > \prod_{n\geq 1} P_n^2\,.
\end{equation}
Here, we used $P_n=P_{-n}$. If this product is finite, then the
survival probability is necessarily finite. We now invoke the
well-known asymptotic behavior \cite{sr},
\begin{equation}
\label{Pn}
1-P_n \sim n^{-(d-2)}\,,
\end{equation}
when $n\gg 1$, to show that the infinite product in \eqref{product}
converges to a finite value when $d>3$. The above analysis can also be 
repeated for a plane, $\delta>2$, and in this case the lower bound is
finite when $d>4$. Hence, the survival probability is finite when the
co-dimension is larger than the critical value, $\Delta > 2$, in
agreement with \eqref{main}.

In summary, we have studied kinetics of a single-species reaction
process, starting with sparse initial conditions.  In our setup,
particles initially occupy a subspace with dimension $d-\Delta$
embedded in a $d$-dimensional space.  The effective dimension of the
diffusion operator equals the co-dimension $\Delta$, and this
parameter controls the asymptotic behavior. The survival probability
of a particle is finite above the critical co-dimension, $\Delta>2$,
but it vanishes otherwise.  We obtained the time-dependent behavior of
the fraction of surviving particles using scaling methods. While our
scaling analysis and numerical simulations were performed for integer
dimensions and co-dimensions, we expect that the results extend to
non-integer dimensions or co-dimensions.  Our investigation complements
studies of catalytic reactions \cite{ob}, and reactions with mobile
and immobile particles \cite{eb} with a similar geometry.

The survival probability \eqref{main} generalizes a classic result in
probability theory. To see this, let us consider two random walkers in
$d$ dimensions, which formally corresponds to the case $\delta=0$.
The separation between these two particles performs a random walk as
well, and the survival probability equals the probability that an
ordinary random walk has yet to visit the origin. Indeed, by
substituting $\Delta=d$ into \eqref{main}, we recover the P\'{o}lya
theorem \cite{polya}.  Hence, our main result \eqref{main} generalizes
the P\'{o}lya theorem from a single random walk to a system of
infinitely many random walkers that interact through the annihilation
process.

There are a number of natural extensions of this work. An immediate
generalization is to aggregation processes. The aggregate mass should
no longer be uniform in space, and it is natural to investigate the
spatial dependence of the mass distribution.  It would also be
interesting to study the two-species annihilation process $A+B\to
\emptyset$ under sparse initial conditions. For this reaction scheme,
starting from homogeneous initial conditions, the critical dimension
is higher, $d=4$ \cite{zo}.  Our study finds that the critical
co-dimension is $\Delta=2$, and hence, it is plausible that there are
multiple regimes of behavior for reactions involving multiple species.

\medskip\noindent We acknowledge support from the US-DOE grant
DE-AC52-06NA25396 (EB) and the BSF Grant No. 2012145 (PLK).

\end{document}